\documentclass[pra,aps,a4paper,showpacs,floatfix,twocolumn,nofootinbib]{revtex4-1}
\usepackage{amsmath}
\usepackage{amsfonts}
\usepackage{amssymb}
\usepackage{graphicx}
\usepackage{dcolumn}
\usepackage{bm}

\DeclareMathOperator{\sinc}{sinc}

\begin{document}

\title{Nonsmooth and level-resolved dynamics illustrated with a periodically driven tight binding model}


\author{J.~M.~Zhang }
\email{wdlang@pks.mpg.de}

\author{Masudul Haque }
\email{haque@pks.mpg.de}

\affiliation{Max Planck Institute for the Physics of Complex Systems, N\"
othnitzer Str.~38, 01187 Dresden, Germany}

\begin{abstract}
We point out that in the first order time-dependent perturbation theory, the transition probability may behave nonsmoothly in time and have kinks periodically. Moreover, the detailed temporal evolution can be sensitive to the exact locations of the eigenvalues in the continuum spectrum, in contrast to coarse-graining ideas. Underlying this nonsmooth and level-resolved dynamics is a simple equality about the sinc function $\sinc x \equiv \sin x / x$. These physical effects appear in many systems with approximately equally spaced spectra, and is also
robust for larger-amplitude coupling beyond the domain of perturbation theory.
We use a one-dimensional  periodically driven tight-binding model to illustrate these effects, both within and outside the perturbative regime. 
\end{abstract}

\maketitle

\section{introduction}

Coarse graining is a common trick in physics. In principle, it is invoked whenever one replaces a summation by an integral. The idea behind, which is justified for many purposes, is that structures too fine do not matter; what matter are the structures on a sufficiently large scale. A good example is calculating the heat capacity of a cloud of ideal gas. The exact single particle eigenvalues of course depend on the specific shape and size of the container. However, for a macroscopic system, one does not need to know the spectrum to that precision at all. On the other hand, one just needs to know 
the number of states in a macroscopic energy interval. In other words, we can replace the real density of states, which consists of delta functions, by a coarse-grained one, which is a continuous and smooth function of energy. 
According to the Weyl's law \cite{weyl}, the coarse-grained density of state is proportional to the volume of the container but independent of its shape. The problem is thus greatly simplified and we have shape independent heat capacity. 

Coarse graining is also used in the definition and derivation of the celebrated Fermi's golden rule \cite{dirac, fermi}. Suppose we have an unperturbed Hamiltonian $\hat{H}_0$, whose eigenvalues and eigenstates are denoted as $E_n$ and $|n\rangle$. Let the system be in the state $|i\rangle $ initially and let a sinusoidal perturbation $\hat{V}(t)= \mathcal{V} e^{i \omega t} + \mathcal{V}^\dagger e^{-i \omega t}$ be turned on at $t=0$. In the first order time-dependent perturbation theory, Fermi's golden rule states that the transition rate from the initial state to the continuum band $\{| n\rangle \}$ is given by ($\hbar = 1$ in this paper) 
\begin{eqnarray}
w_{i\rightarrow [n]} \equiv - \frac{d p }{d t }= 2\pi \overline{|\langle n |\mathcal{V} |i\rangle |^2} \rho (E_n) \big|_{E_n \simeq E_i \pm \omega }.
\end{eqnarray}
Here $p(t)$ denotes the probability of the system remaining in the initial state. 
In this formula, coarse graining is used twice; the quantity $\rho(E_n)$ is the coarse-grained density of states mentioned above while $\overline{|\langle n |\mathcal{V} |i\rangle |^2} $ is the coarse-grained coupling strength. The coarse graining implies that the transition dynamics is smooth with respect to the driving frequency $\omega $.
To be specific, $p(t)$ as a function of $t$ should be largely invariant if the value of $\omega $ is changed on the scale of the level spacing of the spectrum $\{ E_n \} $. 
Or, the transition dynamics cannot resolve the finest structure of the spectrum. 

However, in this paper we point out that while this presumption is true in a finite time interval, it may break down beyond some critical time (the Heisenberg time actually). Specifically, under some mild conditions,  the function $p(t)$ can 
be nonsmooth and have kinks periodically. In the regime where the first order perturbation theory is valid, it is simply a piecewise linear function. Moreover, under the same perturbation, the trajectories of $p(t)$ can bifurcate suddenly and significantly for two adjacent initial states. Similarly, for the same initial state, the trajectory of $p(t)$ can be tuned to a great extent by changing the frequency $\omega $ on the scale of the level spacing. In one word, the transition dynamics can be nonsmooth and have a single level resolution beyond some critical time. These effects can be demonstrated by using the one-dimensional tight-binding model as we shall do below. 

We note that some similar nonsmoothness effect has been observed previously in the model of a single two-level atom interacting with a one-dimensional optical cavity \cite{parker, meystre, stey, ligare}. However, those authors had a different perspective and their approaches were either purely numerical \cite{parker, meystre} or based on an exactly soluble model \cite{stey, ligare}. In contrast, our approach will be based on the first order perturbation theory and some simple mathematical properties of the $\sinc x \equiv \sin x / x $ function. The perturbative approach means some new insight and it enables us to make predictions about a \textit{generic} model, such as the one-dimensional tight-binding model. It also demonstrates that Fermi's golden rule can break down even in the first order perturbation regime ($1-p(t) \ll 1$).

The rest of the paper is organized as follows. In Sec.~\ref{math}, we derive the effect from the first order perturbation theory, assuming that the energy levels in the target region are equally spaced and the couplings to them are equal too. The problem reduces to periodic sampling of the function $\sinc^2 x$, which is to be solved in Sec.~\ref{sincsec} using the Poisson summation formula. There we obtain a piecewise linear function of time, which is the essence of the effect. Then in Sec.~\ref{tbmsec}, we demonstrate 
the effect by taking two examples of tight-binding lattices and driving a local parameter (potential of a single site). In these realistic models, the two assumptions are only approximately satisfied, but we still see sharp kinks (nonsmooth behavior) and sudden bifurcations (level resolution). Moreover, all these effects persist even beyond the perturbative regime, i.e., for large-amplitude driving.

\section{general formalism}\label{math}

Let us recall that in the first order time-dependent perturbation theory, the probability of finding the system in states other than the initial state is given by \cite{decay,sakurai}
\begin{eqnarray}\label{exp1}
1- p(t) = 4\sum_{n \neq i }  |\langle n |\mathcal{V}|i \rangle |^2   \frac{  \sin^2 [(E_n-E_f )t/2]}{|E_n - E_f|^2} ,
\end{eqnarray}
with $E_f \equiv  E_i + \omega $.
Here we assume that stimulated absorption is the only dominant process. The case when stimulated absorption and stimulated radiation are dominant simultaneously can be treated equally well.
Now we make two assumptions: (i) the level spacing $E_{n+1}- E_n $ and (ii) the coupling strength $|\langle n |\mathcal{V}|i \rangle | $ are both slowly varying with respect to $n$ for $E_n \simeq E_f $. We will discuss later for what kinds of model these assumptions are satisfied. 

Under these assumptions, in view of the fact that the $\sinc^2 x$ function decays in the rate of $|x|^{-2}$ to zero as $|x| \rightarrow \infty$, it is a good approximation to replace the real spectrum in (\ref{exp1}) by an equally spaced spectrum and the real couplings by a constant one. That is, 
\begin{eqnarray}\label{firstappro}
1- p (t) \simeq 4 g^2 \sum_{m\in \mathbb{Z}} \frac{  \sin^2 [(\tilde{E}_m-E_f )t/2]}{|\tilde{E}_m - E_f|^2} .
\end{eqnarray}
Here the pseudo spectrum is defined as $\tilde{E}_m = E_{n^*}+ m (E_{n^* +1}-E_{n^*})$
, $m\in \mathbb{Z}$, and the constant coupling is $g = |\langle n^* |\mathcal{V}|i \rangle |$. The level $|n^*\rangle $ is chosen by the condition $E_{n^*} \leq  E_f < E_{n^*+1}$. 
Defining $\delta = E_{n^* +1}-E_{n^*} $ and $ \alpha = (E_f - E_{n^*})/\delta$, we can rewritten (\ref{firstappro}) as 
\begin{eqnarray}\label{w1}
1- p (t) \simeq \left( \frac{4 g^2}{\delta^2 } \right) W_\alpha  (T)
\end{eqnarray}
Here, we have introduced the function of the rescaled dimensionless time $T = \delta t/2$,
\begin{eqnarray}\label{w2}
W_\alpha (T) \equiv  T^2 \sum_{m \in \mathbb{Z}} \sinc^2 [(m-\alpha )T],
\end{eqnarray}
which will be our primary concern. The summation means to sample the function $\sinc^2 x $ uniformly from $-\infty $
to $+ \infty$ with an equal distance $T$. The offset is determined by $\alpha$. Apparently, $W_\alpha = W_{\alpha+1}$ and because of the evenness of the $\sinc $ function, $W_\alpha = W_{-\alpha }$. Therefore, $W_\alpha $ is determined by its value in the interval of $0\leq \alpha \leq 1/2$.

\begin{figure*}[tb]
\includegraphics[width= 0.4\textwidth ]{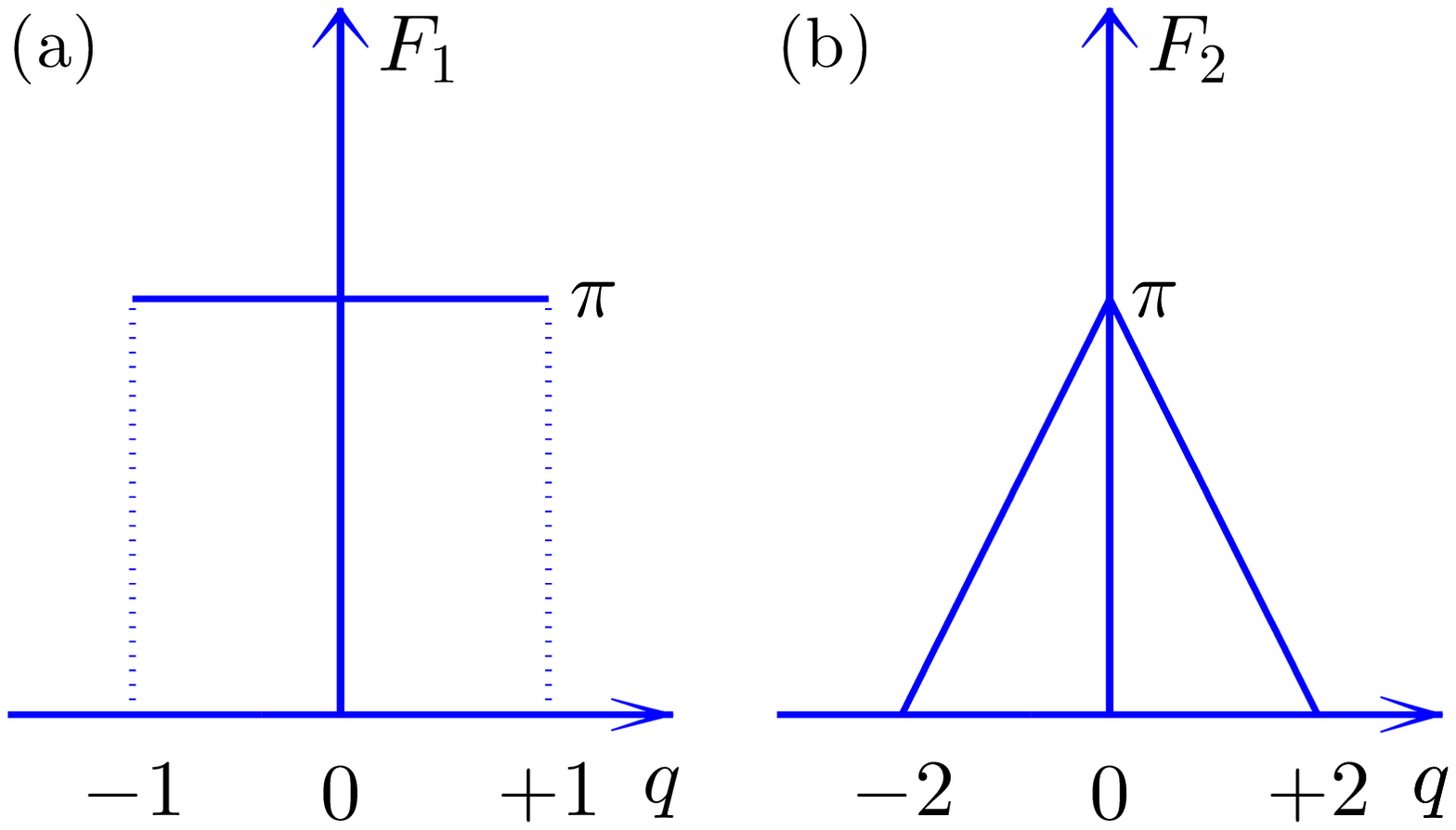}\hspace{20pt}
\includegraphics[width= 0.385\textwidth ]{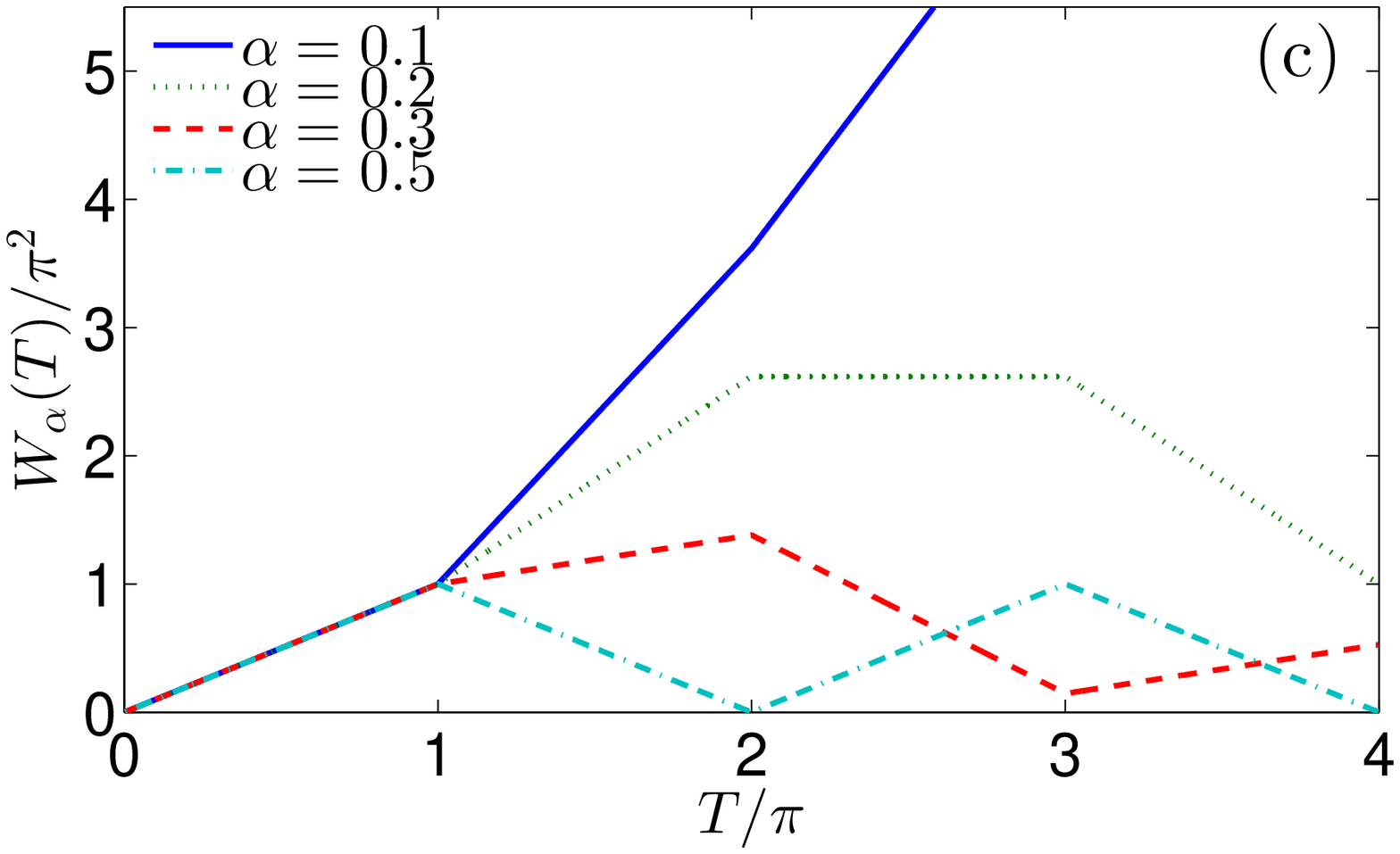}
\caption{(Color online) (a) Fourier transform of the function $\sinc x$ and (b) Fourier transform of its square $\sinc^2 x$. The latter is calculated from the former by convolution. The point is that both have a finite support. In (c), the function $W_\alpha (T)$ [see Eqs.~(\ref{w2}) and (\ref{wfull})], which is calculated by using the Poisson summation formula and the fact in (b), is displayed for four different values of the offset $\alpha $. It is piecewise linear. In the interval of $0\leq T \leq \pi$, its value is independent of the parameter $\alpha $; while beyond this interval, its value is very sensitive to $\alpha $. Note that for $\alpha = 0.5$, $W_\alpha $ returns to zero at $T= 2 m \pi $ periodically; while for $\alpha = 0$, $W_\alpha \propto T^2$ asymptotically. }
\label{fourier}
\end{figure*}

\subsection{An equality of the  $\sinc^2 x$ function}\label{sincsec}

To carry out the summation in (\ref{w2}), we employ the celebrated Poisson summation formula \cite{grafakos}, which essentially means that periodically sampling a function in real space is equivalent to periodically sampling (with appropriate phase shifts) the same function in momentum space. To this end, we need the Fourier transform of $\sinc^2 x$. Because it is the square of the function $\sinc x$, in turn, we need the Fourier transform of $\sinc x$, from which the Fourier transform of $\sinc^2 x$ can be calculated by convolution. But it is well known that $\sinc $ is the Fourier transform of the window function! That is, the function defined as 
\begin{eqnarray}
F_1 (q ) \equiv \int_{-\infty}^{+\infty } d x e^{-i q x } \frac{\sin x}{ x},
\end{eqnarray}
which can be calculated by using complex contour integration, is 
\begin{eqnarray}
F_1 (q ) = \begin{cases} \pi, & |q|\leq 1, \\ 0, & |q| > 1.  \end{cases}
\end{eqnarray} 
It has a finite support of $[-1, + 1 ]$ and has a constant value of $\pi$ there, as illustrated in Fig.~\ref{fourier}(a). By convolution, the Fourier transform of $\sinc^2 x$ is then 
\begin{eqnarray}
F_2(q ) = \begin{cases} \frac{\pi}{2} (2+q) , & -2\leq q \leq 0 , \\ \frac{\pi}{2} (2-q) , & 0\leq q \leq + 2  , \\ 0, & |q| > 2 , \end{cases} 
\end{eqnarray}
which is a triangle function on the support $[-2, +2]$ as illustrated in Fig.~\ref{fourier}(b). We note that the fact that $\sinc ^2 x$ has
a finite support in the Fourier space is a consequence of the
Paley-Wiener theorem \cite{rudin}. The function is entire and is of
exponential type 2. Therefore, by the Paley-Wiener theorem,
its Fourier transform is supported on $[-2, +2]$.

Now by the Poisson summation formula, we have 
\begin{eqnarray}\label{w3}
W_\alpha (T) & = & T \sum_{n\in \mathbb{Z}} F_2 \left(\frac{2\pi n }{T} \right ) \exp \left(-i 2\pi n  \alpha \right).
\end{eqnarray}
We have thus successfully converted the original, equally spaced sampling of $\sinc^2 x$, into an equally spaced sampling of its Fourier transform $F_2(q)$. Now the point is that, this function has only a \textit{finite} support. This means for given $T$, only a finite number of terms on the right hand side of (\ref{w3}) will contribute. In particular, if $0<T\leq \pi$, only the $n=0$ term is nonzero and we get $ W_\alpha (T) = \pi T $, which is simply linearly proportional to $T$. Moreover, it is \textit{independent} of the parameter $\alpha$. By (\ref{w1}), we get $ 1- p (t) \simeq (2\pi g^2/\delta  ) t $
for $0\leq t \leq t_c \equiv 2\pi/\delta$. Here we note that the critical time $t_c$ is the so-called Heisenberg time. This is nothing but Fermi's golden rule if one notes that $1/\delta $ is the coarse-grained density of states.\footnote{Pedagogically, this approach of deriving Fermi's golden rule, although not as general as, has some advantage over the conventional one in textbooks like ref.~\cite{sakurai}, in the sense that the Dirac delta function is not invoked.} 
The $\alpha$-independence also justifies the coarse-graining usually employed. 

However, more generally, if $m \pi < T \leq (m+1) \pi $, the nonzero terms in the summation are $-m \leq n \leq m$. We have then ($\theta \equiv  2\pi \alpha $)
\begin{eqnarray}\label{wfull}
W_\alpha (T) 
&=&  \pi T \sum_{n=-m}^{m} \exp (i n \theta )  - \sum_{n=1}^m  2 \pi^2 n \cos ( n \theta). \quad 
\end{eqnarray}
This is our central result. The function $W_\alpha$
is still linear on the interval $[m \pi, (m+1)\pi ]$, but now the slope \textit{depends} on $\alpha$ and $m $, which means it is a piecewise linear, nonsmooth function of $T$ and has kinks at $T = m \pi $ periodically. In Fig.~\ref{fourier}(c), the function $W_\alpha(T)$ is plotted for four different values of $\alpha$. We see that although on the interval $(0,\pi]$, the lines all coincide, immediately after the first kink, they all split out and their late developments differ significantly. Therefore, we see that even in the first order perturbation theory, Fermi's golden rule can break down beyond some critical time.

At this point, it should be clear why we need the two assumptions at the beginning. We need to sample the $\sinc^2 x$ function uniformly and with equal weight to make use of the nice expression (\ref{wfull}). We argue that the two assumptions are satisfied for a generic model with only one degree of freedom, but are dissatisfied for a generic model with more than one degree of freedom. The reason is simply that if one adds up two arithmetic sequences with different common differences, one does not get an arithmetic sequence. Therefore, we have to admit that, the effects discussed in this paper are relevant only to single particle models in one dimension. 

By (\ref{w1}), the nonsmooth, $\alpha$-dependent dynamics of $W_\alpha$ translates into that of $p(t)$. Both the coupling $g$ and the level spacing $\delta $ are slowly varying functions of $E_i$ and $\omega $. More specifically, they changes only negligibly if $E_i$ or $\omega $ changes on the order of $\delta $. However, $\alpha $ varies on the order of $\delta $. Therefore, we expect that under the same sinusoidal perturbation, adjacent eigenstates generally will have significantly different transition dynamics beyond the critical time $t_c$. Or, if we start from the same initial state but drive with slightly different $\omega $, the $t>t_c$ dynamics has level resolution---it depends on the exact location of $E_f $ 
relative to the eigenvalues. 

\section{Driven tight-binding model}\label{tbmsec}

\begin{figure}[tb]
\includegraphics[width= 0.4\textwidth ]{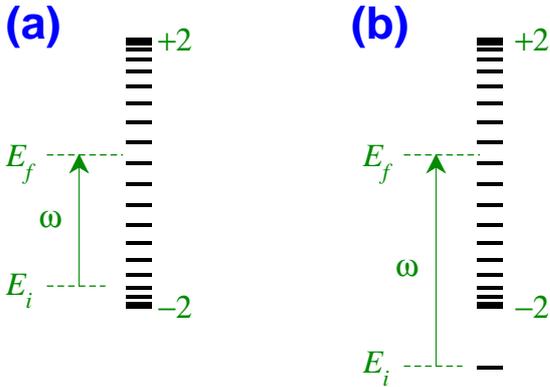}
\caption{(Color online) Two scenarios of transition in the periodically driven one-dimensional tight-binding model. (a) Continuum-to-continuum transition. The initial state is an eigenstate, i.e., a Bloch state, of the perfect model (\ref{h1}). (b) Bound-to-continuum transition. The unperturbed Hamiltonian (\ref{h2}) contains a defected site and the initial state is the defect mode. Note that locally speaking, the spectrum of the tight-binding model is almost equally spaced. }
\label{cartoon}
\end{figure}

\begin{figure*}[tb]
\includegraphics[width= 0.95\textwidth ]{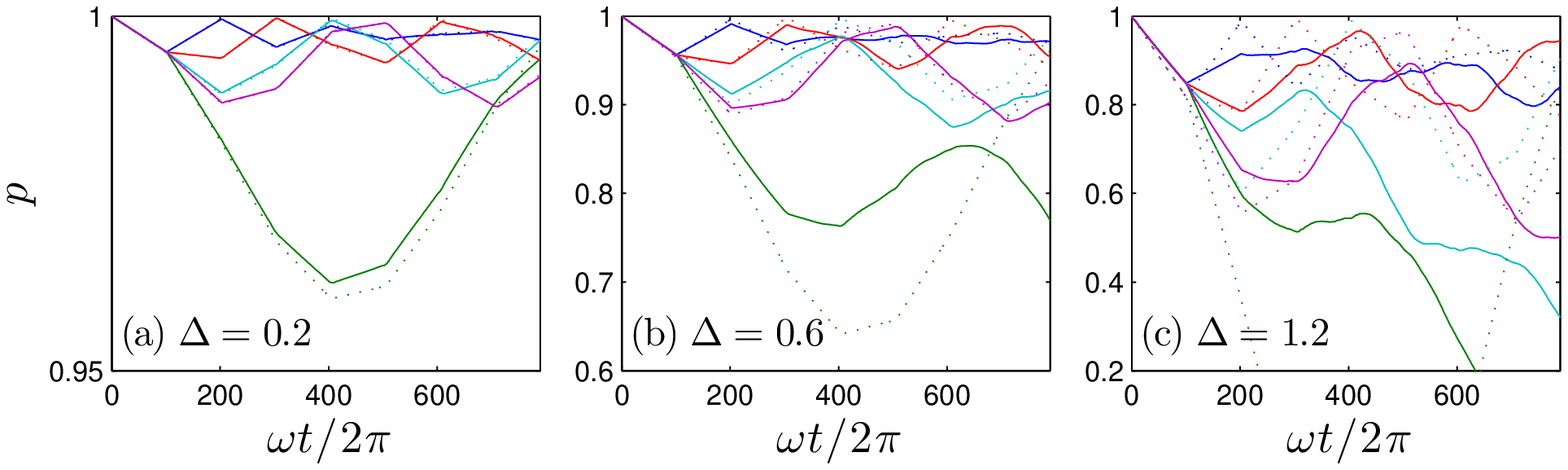}
\caption{(Color online) Probability of finding the particle in the initial Bloch state against the number of driving cycles. The lattice size is $N = 601$ and the period of driving is $2\pi/\omega =3$. The amplitude of driving is shown in each panel. In each panel, five successive initial states, with $41 \leq k_i \leq 45 $ are studied. The solid lines are exact results obtained by solving the time-dependent Schr\"odinger equation, while the dotted lines are based on (\ref{w1}) and (\ref{wfull}). In each panel, at $\omega t/2\pi=200$, from up to down, the solid lines correspond to $k_i =41$, $43$, $44$, $45$, $42 $, respectively. The same ordering holds for the dotted lines. Note that the distance between the kinks is independent of $\Delta$. }
\label{perfect}
\end{figure*}

In the following, we take the one-dimensional tight-binding model to illustrate these effects. Consider a one-dimensional lattice with $N= 2L+1 $ sites and with periodic boundary condition. The Hamiltonian is 
\begin{eqnarray}\label{h1}
\hat{H}_{TBM} = -\sum_{j=-L}^{L} (\hat{a}_j^\dagger \hat{a}_{j+1} + \hat{a}_{j+1}^\dagger \hat{a}_{j}).
\end{eqnarray}
The eigenstates are labelled by the integer $-L \leq k \leq L $. The $k$th eigenstate is $\psi_k (n)=  e^{i 2\pi k n /N} /\sqrt{N }$  and the corresponding  eigenvalue is $\epsilon_k = - 2 \cos (2\pi  k/N) $. Suppose initially the particle is in the state $\psi_{k_i}$ and at $t=0$ we start modulating the potential of site $j=0$ sinusoidally by adding to the Hamiltonian (\ref{h1}) the term
\begin{eqnarray}\label{perturb}
\hat{V}(t) = \Delta  \hat{a}_0^\dagger \hat{a}_0   \sin  \omega t .
\end{eqnarray}
The perturbation couples all the eigenstates with equal amplitude, i.e., $|\langle \psi_{k_2}|\mathcal{V}|\psi_{k_1}\rangle |= \Delta/2N $ for all $k_{1,2}$ \cite{diff}. We choose $k_i$ such that the initial energy $E_i = \epsilon_{k_i}$ is close enough to the bottom of the energy band and we need only to consider the stimulated absorption process [see Fig.~\ref{cartoon}(a)]. 

The two assumptions about the level spacings and couplings are satisfied. 
Let $E_f = E_i + \omega $ fall between $\epsilon_{k_f}$ and $\epsilon_{k_f+1}$ ($k_f >0$). We then approximate the real spectrum by the spectrum $\{ \epsilon_{k_f} + n (\epsilon_{k_f+1} - \epsilon_{k_f})| n \in \mathbb{Z} \}$. We have thus for the parameters $\delta$ and $g$: $\delta =\epsilon_{k_f+1} - \epsilon_{k_f} $ and $g = \Delta/2N $. The parameter $\alpha $ is then determined as $\alpha  = (E_f - \epsilon_{k_f} )/(\epsilon_{k_f+1} - \epsilon_{k_f})$. Both $g$ and $\delta$ are slowly varying with respect to the index $k_i$. However, $\alpha $ is expected to be somewhat random, as demonstrated below. 
We can then use (\ref{w1}) and (\ref{wfull}) to predict the time evolution of $p$ [but here we have to multiply the right hand side of (\ref{wfull}) by a factor of two because a generic level of (\ref{h1}) is doubly degenerate] to the first order of perturbation. 

In Fig.~\ref{perfect}, we compare the approximate results obtained in this way with the results obtained by solving the time dependent Schr\"odinger equation exactly. 
There, we have taken a lattice of $N=601$ sites and studied the transition dynamics of five successive plane waves, i.e., for $41 \leq k_i \leq 45 $, under the same driving. 
We see that when the driving amplitude $\Delta$ is small enough [Fig.~\ref{perfect}(a)], the exact evolution of $p$ agrees with the approximation very well. The trajectories of $p$ show kinks periodically and between the kinks, they are all linear. The trajectories all collapse into one before the first kink. However, immediately after the first kink, they spread out, somewhat randomly in the sense that they are not ordered as their wave vectors. As the driving amplitude increases [Figs.~\ref{perfect}(b) and \ref{perfect}(c)], the exact result deviates from the approximate prediction gradually, and the kinks get rounded gradually, starting first with the later ones. However, even when the first order perturbation is no longer good quantitatively in the regime $0< t < t_c$ [see Fig.~\ref{perfect}(c)], the first kink still happens and we still observe the state dependent spread-out. Moreover, the second kink is still visible for those trajectories with $p \lesssim 1$ marginally satisfied.  

As a second scenario, let us consider a one-site-defected tight-binding model. Now the Hamiltonian is ($U>0$)
\begin{eqnarray}\label{h2}
\hat{H}_{TBM}' = -\sum_{j=-L}^{L-1} (\hat{a}_j^\dagger \hat{a}_{j+1} + \hat{a}_{j+1}^\dagger \hat{a}_{j}) - U \hat{a}_0^\dagger \hat{a}_{0}.
\end{eqnarray}
The defect introduces a localized mode $\phi_d$ around it. For a sufficiently big $L$, its energy is $E_i = -\sqrt{U^2 + 4}$. Suppose initially the particle is trapped in this mode. Under the perturbation (\ref{perturb}), it couples to the eigenstates in the continuum band [illustrated in Fig.~\ref{cartoon}(b)]. Note that the model (\ref{h2}) is symmetric with respect to the defected site and therefore all its eigenstates have a definite parity. In particular, the defect mode has an even parity. The perturbation (\ref{perturb}) preserves the parity symmetry and therefore, it couples the defect mode only to those even-parity states. 

Denote the even-parity eigenstates as $\phi_m$ and correspondingly, the eigenvalues as $\{ \varepsilon_m \}$. Numerically, it is verified that the level spacing $\varepsilon_{m+1} - \varepsilon_m $, as well as the coupling $|\langle \phi_m | \mathcal{V}| \phi_d \rangle |$, is slowly varying with respect to $m $. Let $E_f = E_i + \omega $ fall between $\varepsilon_n $ and $\varepsilon_{n+1}$. As in the first example, we approximate the real spectrum by $\{ \varepsilon_n + m (\varepsilon_{n+1} - \varepsilon_n ) | m \in \mathbb{Z} \}$, which means $\delta = \varepsilon_{n+1} - \varepsilon_n $, and take $g$ as $g = |\langle \phi_n | \mathcal{V}| \phi_d \rangle |$. 
Both $\delta $ and $g$ are almost invariant when $\omega $ changes on the scale of $\delta$. However, the parameter $\alpha  = (E_f - \varepsilon_{n} )/\delta $ can be changed significantly in this process.

\begin{figure}[tb]
\includegraphics[width= 0.45\textwidth ]{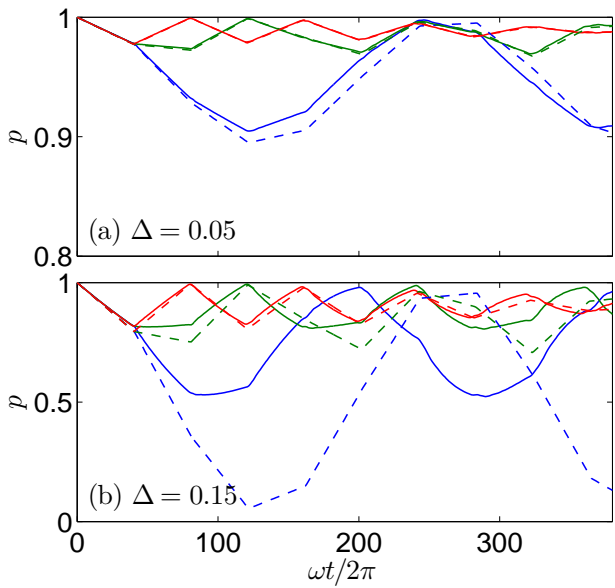}
\caption{(Color online) Probability of finding the particle in the initial state (the defect mode) against the number of driving cycles. The lattice size is $N = 201$ and the depth of the defect is $U=1$. The amplitude of driving is shown in each panel. In each panel, the solid lines are exact results obtained by solving the time-dependent Schr\"odinger equation, while the dashed lines are based on (\ref{w1}) and (\ref{wfull}). In each panel, at $\omega t/2\pi =100$, from bottom to top, the solid lines correspond to $2\pi/\omega = 2.5$, $2.51$, $2.52$, respectively. The same ordering holds for the dashed lines. Note that the distance between the kinks is independent of $\Delta$. }
\label{local}
\end{figure}

Therefore, the transition dynamics might change substantially as $\omega $ changes on the scale of the mean level spacing of the model. This is indeed the case, as shown in Fig.~\ref{local}. There, a relatively weak [Fig.~\ref{local}(a)] and a relatively strong driving case [Fig.~\ref{local}(b)] are studied. In both cases, three slightly different driving periods are considered. Like in Fig.~\ref{perfect}, we see periodic kinks and the spread-out of the trajectories 
at the first kink. In the stronger driving case, the exact trajectories no longer follow the predictions based on the approximation (they simply cannot, as the approximation predicts $p\ll 1$ at some point, which is inconsistent with the perturbation theory). The trajectories even look curved. However, the periodic kinks remain as the most apparent features. 

So far, our discussion is based on mathematics only. Here are some remarks from a physical point of view. First, the level resolution dynamics belongs to the long-time regime $t>t_c= 2\pi /\delta $. This is consistent with the usual time-energy uncertainty relation---one has to wait a time on the order of $1/\delta $ to have an energy resolution of $\delta $. Second, the periodic kinks are a consequence of interference. The local driving generates some out-going waves. These waves are peaked in momentum space and travel along the lattice with a group velocity of
\begin{eqnarray}
v_g \equiv  \frac{d E}{d k } = \frac{\delta }{(2\pi/N)} = \frac{N \delta}{2 \pi }.
\end{eqnarray} 
In time $N/v_g$, which equals $ t_c $, they come back to the perturbed site and interfere with the newly generated out-going waves. This picture explains why in Figs.~\ref{perfect} and \ref{local} the kinks are robust even beyond the perturbative regime. 

\section{Conclusion and discussion}

In conclusion, we have reexamined the first order time-dependent perturbation theory. 
By noticing some simple analytic equality about the $\sinc $ function, we predicted that under some mild conditions, the probability of finding the system in its initial state 
is a piecewise linear function of time. The slope of the line changes periodically with the period being inversely proportional to the level spacing of the spectrum. 
These predictions were confirmed in the one-dimensional tight-binding model. 
Although the predictions are based on the first order time-dependent perturbation theory, the signatures persist even outside of the perturbative regime. 

It should be worthwhile to construct some system to observe the  predicted effects experimentally. A potential scheme is to use guiding photonic structures \cite{longhi}. By the quantum-optics analogy, the Hamiltonians (\ref{h1}) and (\ref{h2}), as well as the time-varying driving (\ref{perturb}), can all be readily realized. The driving does not need to be sinusoidal actually, it can equally well be in the periodic square form, which is experimentally  more feasible. The main difficulty might come from the relatively large number of sites required for a clean manifestation of the effects. 

\section*{Acknowledgments}

We are grateful to Andr\'e Eckardt, Hua-Tong Yang, Hui Jing, Michael Sekania, and Alex D. Gottlieb for helpful discussions.


\begin{thebibliography}{99}
\bibitem{decay}
Although here in view of the demonstration in Sec.~\ref{tbmsec}, we have assumed a sinusoidal perturbation, the formalism applies  equally well to the case of a constant perturbation.
\bibitem{diff}
Here the perturbation couples not only the initial state and the continuum states, but also the continuum states themselves. In contrast, in the exactly soluble model in Refs.~\cite{meystre,stey,ligare}, there is no coupling between the continuum states. Of course, in the first order perturbation, the coupling between the continuum states does not appear.
\end{thebibliography}
\end{document}